\documentclass[12pt]{article}

\usepackage{graphicx}
\usepackage{etoolbox}
\usepackage{hyperref}
\usepackage{times}
\usepackage{bm}
\usepackage{makecell}
\usepackage{color}
\usepackage{xcolor}
\usepackage[numbers,sort&compress]{natbib}

\usepackage{amssymb}
\usepackage{amsmath}
\usepackage{braket}
\usepackage{siunitx}
\usepackage{lineno}
\usepackage[normalem]{ulem}
\newenvironment{sciabstract}{%
\begin{quote} \bf}
{\end{quote}}

\graphicspath{{Figures/}}

\makeatletter
\let\saved@includegraphics\includegraphics
\AtBeginDocument{\let\includegraphics\saved@includegraphics}

\makeatother

\DeclareSIUnit\gauss{G}

\begin{document}

\title{Colossal low-field negative magnetoresistance in CaAl$_2$Si$_2$-type diluted magnetic semiconductors (Ba,K)(Cd,Mn)$_2$As$_2$}


\date{}

\author{%
B.~J.~Chen$^{1,\ast}$,
Z.~Deng$^{1}$,
C.~Q.~Jin$^{1,2,\ast}$\\[0.6em]
\parbox{\textwidth}{\centering\small
$^{1}$ Beijing National Laboratory for Condensed Matter Physics, Institute of Physics, Chinese Academy of Sciences, Beijing 100190, China\\
$^{2}$ Collaborative Innovation Center of Quantum Matter, Beijing 100190, China\\
$^{\ast}$ To whom correspondence should be addressed.  E-mail: bijuanchen.5459@gmail.com; Jin@iphy.ac.cn
}}
\date{}

\baselineskip24pt
\maketitle

\vspace{5mm}

\begin{sciabstract}

We report the magnetic and magnetotransport properties of the layered CaAl$_2$Si$_2$-type diluted magnetic semiconductor (Ba$_{1-x}$K$_x$)(Cd$_{1-y}$Mn$_y$)$_2$As$_2$ over a broad Mn (spin) substitution range of 0.05 $\le$ $y$ $\le$ 0.5. K substitution introduces hole carriers, whereas Mn provides local moments, resulting in bulk ferromagnetism with Curie temperatures up to $\sim$ 17~K. Intrinsic magnetic ordering is further supported by an anomalous Hall contribution and a specific-heat anomaly near $T_{\mathrm{C}}$. A key performance feature is a colossal negative magnetoresistance: for heavily Mn-doped compositions ($y$ $\ge$ 0.3), $MR=[\rho(H)-\rho(0)]/\rho(0)$ reaches approximately $-100\%$ at 2~K and nearly saturates at a relatively low magnetic field of $\sim$ 0.35~T. The combination of soft ferromagnetism, strong spin--charge coupling, and low-field MR saturation highlights (Ba,K)(Cd,Mn)$_2$As$_2$ as a promising bulk platform for low-temperature magnetoresistive functionalities.

\end{sciabstract}

\subsection*{Introduction}

Magnetoresistive materials underpin semiconductor spintronics by enabling electrical readout and manipulation of magnetic states, which are essential for memory, sensing, and spin-based information processing~\cite{ref1_awschalom2007,ref2_zutic2004}.
Within this space, diluted magnetic semiconductors (DMSs) are particularly attractive because they combine semiconducting transport with tunable magnetism, providing a chemically flexible route to integrate magnetic and electronic functionalities in a single materials platform~\cite{ref4_dietl2010}.
From a materials perspective, a central objective is to realize large magnetoresistance ($MR$) at low magnetic fields while maintaining robust magnetic order and controllable carrier density, since these metrics directly impact switching efficiency and device-readout margins.

Early DMS research was driven by Mn-doped III--V semiconductors, especially (Ga,Mn)As, which established the concept of ferromagnetism emerging in a semiconductor host~\cite{ref3_ohno1998,ref5_ohno1996,ref8_jungwirth2006,ref34_matsukura1998}.
Subsequent materials and nanostructure engineering pushed the Curie temperature of (Ga,Mn)As toward $\sim$200~K, underscoring both the promise and the practical challenges of this platform~\cite{ref13_wang2008,ref12_chen2011}.
At the same time, these systems illustrate a key limitation: the magnetic dopant can simultaneously introduce both local moments and carriers, constraining solubility and complicating independent control of spin and charge~\cite{ref4_dietl2010,ref6_samarth2010,ref7_chambers2010}.
This motivates the \emph{decoupled spin and charge doping} strategy in bulk materials, where one substitution channel primarily tunes carrier density while another supplies magnetic moments, enabling clearer structure--chemistry--property relationships and more systematic materials optimization.

Bulk DMS families implementing decoupled doping have expanded rapidly. A representative example is the I--II--V host LiZnAs, where Mn substitution provides local moments and carriers can be tuned independently, establishing a robust bulk DMS framework~\cite{ref9_deng2011,ref30_deng2013}.
In parallel, pnictide-derived bulk systems provide broader chemical design space and higher ordering temperatures, including the ``122''-type (ThCr$_2$Si$_2$-structure) DMS (Ba,K)(Zn,Mn)$_2$As$_2$~\cite{ref10_zhao2013,ref11_zhao2014csb}.
Related codoping strategies, such as Mn/Cu substitution at the Zn sites in Ba(Zn$_{1-2x}$Mn$_x$Cu$_x$)$_2$As$_2$, further demonstrate the compositional tunability of bulk DMS platforms~\cite{ref19_man2015}.
Moreover, spectroscopic probes have clarified carrier--moment coupling in the ``122'' family, providing a basis for linking materials chemistry to electronic structure and magnetic exchange~\cite{ref21_suzuki2015prb,ref24_suzuki2015prb_arpes}.

Beyond maximizing $T_{\mathrm{C}}$, an increasingly active direction is to engineer extreme magnetotransport performance, particularly large negative $MR$.
Layered pnictide DMS systems adopting the hexagonal CaAl$_2$Si$_2$-type structure are a useful platform because they enable decoupled doping within a layered framework and often exhibit semiconducting or localization-influenced transport, which can amplify field-tunable resistivity changes.
Representative CaAl$_2$Si$_2$-type DMS materials include (Sr,Na)(Zn,Mn)$_2$As$_2$ and (Ca,Na)(Zn,Mn)$_2$As$_2$, and extend to Cd-based analogs such as (Sr$_{1-x}$Na$_x$)(Cd$_{1-x}$Mn$_x$)$_2$As$_2$, establishing broad chemical accessibility of this structure type~\cite{ref14_chen2014prb,ref15_zhao2014jap,ref17_chen2016jap}.
Notably, colossal negative $MR$ has been reported in the related CaAl$_2$Si$_2$-type system (Sr,K)(Zn,Mn)$_2$As$_2$, highlighting the potential for large $MR$ responses in bulk DMS materials with chemically tunable disorder and magnetic exchange~\cite{ref18_yang2014epl}.
Follow-up work further discussed magnetic and transport mechanisms in this system~\cite{ref22_yang2016jmmm}.
More recent efforts continue to expand the magnetotransport landscape through approaches such as chemical-pressure tuning in bulk DMS systems~\cite{ref40_yu2019aplmater},
extension to additional CaAl$_2$Si$_2$-type DMS variants~\cite{ref41_dong2022jsemicond},
and reports of very large negative $MR$ in other decoupled-doping or disorder-tuned magnetic semiconductors (e.g., Na(Zn,Mn)Sb)~\cite{ref42_yu2023jsemicond,ref44_zhao2025prb}.
In addition, emerging layered DMS single-crystal studies highlight the importance of separating intrinsic transport from microstructure-driven effects when benchmarking $MR$ performance~\cite{ref43_peng2024nanomaterials}.

Within this context, (Ba,K)(Cd,Mn)$_2$As$_2$ is a compelling CaAl$_2$Si$_2$-type platform in which K substitution introduces hole carriers and Mn substitution provides local moments, enabling systematic tuning of spin--charge coupling in a layered arsenide host.
Earlier work established this compound family as a bulk DMS with sizable negative $MR$~\cite{ref16_yang2013jap}.
Here we extend the compositional window to a broad Mn substitution range (0.05$\le$ $y$ $\le$0.5) and emphasize the high-$y$ regime where $MR$ performance is maximized. Compared with the earlier report on this system (Ref.~\cite{ref16_yang2013jap}), which covered $y \leq 0.2$ and reported $MR$ up to approximately $-70\%$, the present work reveals a regime of colossal negative $MR$ approaching $-100\%$ with low-field saturation at $\sim 0.35\,\mathrm{T}$, and provides additional thermodynamic and Hall-effect evidence supporting magnetic ordering not reported previously.
By combining magnetization, Hall, and heat-capacity measurements, we strengthen the case for intrinsic magnetic ordering (including an anomalous Hall contribution and a specific-heat anomaly near $T_{\mathrm{C}}$), and we highlight magnetotransport performance in heavily Mn-doped compositions, which exhibit colossal negative $MR$ approaching $-100\%$ at low temperature with near-saturation at relatively low magnetic fields.
Together, these results position (Ba,K)(Cd,Mn)$_2$As$_2$ as a chemically tunable bulk DMS platform for low-temperature magnetoresistive functionalities and provide materials-level guidance for optimizing $MR$ through composition and magnetic disorder.

\subsection*{Experimental methods}

\emph{Sample synthesis and Structural characterization}

Polycrystalline (Ba$_{1-x}$K$_x$)(Cd$_{1-y}$Mn$_y$)$_2$As$_2$ samples were prepared by a solid-state reaction method. We note that all compositions reported here are nominal values based on the precursor ratios. The intended substitution is supported by the phase-pure XRD patterns and the systematic evolution of lattice parameters with K and Mn doping (Fig.~1c). Two composition series were synthesized: (i) a K-doping series (Ba$_{1-x}$K$_x$)(Cd$_{0.9}$Mn$_{0.1}$)$_2$As$_2$ with $x$=0.01, 0.03, 0.04, 0.05, 0.07, and 0.10, and (ii) a Mn-doping series (Ba$_{0.96}$K$_{0.04}$)(Cd$_{1-y}$Mn$_y$)$_2$As$_2$ with 0.05$\le$ $y$ $\le$0.5. Starting materials (nominal compositions), including potassium arsenide, barium arsenide, arsenic pieces, cadmium powder, and manganese powder, were thoroughly mixed, loaded into tantalum tubes, sealed under Ar, and heated at 700~$^\circ$C for 24~h, followed by furnace cooling to room temperature. The reacted products were reground, pelletized, sealed in evacuated tantalum tubes, and annealed at 500~$^\circ$C for 24~h (following a procedure similar to that used for BaFZnAs)~\cite{ref37_chen2016cpb}.

Room-temperature powder X-ray diffraction (XRD) was performed using a Philips X'Pert diffractometer with Cu K$\alpha$ radiation. Rietveld refinements were carried out using the GSAS software package~\cite{ref38_larson1994gsas}.

\emph{Magnetic, transport, and thermodynamic measurements}

DC magnetization measurements were performed using a Quantum Design SQUID-VSM in the temperature range 2--300~K. Electrical transport measurements were carried out using a Quantum Design PPMS in the temperature range 2--300~K with a standard four-probe configuration; electrical contacts were made using silver paste. Magnetoresistance and Hall measurements were performed in the same PPMS system under applied magnetic fields. Specific heat was measured using the heat-capacity option of the Quantum Design PPMS in the temperature range 2--300~K.

\subsection*{Results and discussion}

\emph{Crystal structure and composition evolution}

Figure~1b shows the powder X-ray diffraction (XRD) patterns of (Ba$_{1-x}$K$_x$)(Cd$_{0.9}$Mn$_{0.1}$)$_2$As$_2$ for $x$=0.01, 0.03, 0.04, 0.05, 0.07, and 0.10. No structural transition is observed across this composition range. All reflections can be indexed using the hexagonal CaAl$_2$Si$_2$-type structure with space group $P\bar{3}m1$ (Fig.~1a), consistent with the parent compound BaCd$_2$As$_2$~\cite{ref25_klufers1984,ref26_mewis1980}.
In BaCd$_2$As$_2$, Ba occupies the one-fold Wyckoff position $1a$ (0,0,0), while Cd and As occupy the two-fold Wyckoff position $2d$ (1/3, 2/3, $z$).
The structure is layered along the crystallographic $c$-axis, consisting of Cd$_2$As$_2$ layers separated by Ba/K layers. The Cd-centered [CdAs$_4$] tetrahedra form the Cd$_2$As$_2$ layers, while the Ba/K atoms occupy [BaAs$_6$] octahedral sites between adjacent Cd$_2$As$_2$ layers. Viewed along the $c$-axis, the Cd$_2$As$_2$ layers exhibit a corrugated honeycomb-like network (Fig.~1a).
This layered hexagonal framework is distinct from the tetragonal ``122''-type ferromagnetic DMS family (Ba,K)(Zn,Mn)$_2$As$_2$~\cite{ref10_zhao2013,ref11_zhao2014csb}.

Lattice parameters refined using the GSAS software package~\cite{ref38_larson1994gsas} are summarized in Fig.~1c.
Compared with the parent BaCd$_2$As$_2$ ($a$=4.4919~\AA\ and $c$=7.6886~\AA), both $a$ and $c$ expand with K doping, whereas they contract with Mn doping (inset of Fig.~1c), consistent with the intended chemical substitution.
Such systematic lattice evolution is consistent with trends reported for related CaAl$_2$Si$_2$-type bulk DMS families~\cite{ref14_chen2014prb,ref15_zhao2014jap}.

\emph{K-doping dependence of ferromagnetism}

To separate the effects of charge and spin doping, we vary one dopant concentration while fixing the other. Figure~2a shows the temperature dependence of magnetization $M(T)$ measured under 500~Oe for (Ba$_{1-x}$K$_x$)(Cd$_{0.9}$Mn$_{0.1}$)$_2$As$_2$ with $x$=0.01, 0.03, 0.04, 0.05, 0.07, and 0.10, revealing clear signatures of ferromagnetic ordering. No discernible difference is observed between the zero-field-cooled (ZFC) and field-cooled (FC) magnetization curves (see Fig.~S3 in the Supporting Information), indicating the absence of pronounced magnetic irreversibility or canonical spin-glass-like freezing behavior in this system.
Above $T_{\mathrm{C}}$, the samples are paramagnetic. In the paramagnetic regime, the susceptibility can be described by the Curie--Weiss form (inset of Fig.~2a):
\begin{equation}
(\chi - \chi_0)^{-1} = \frac{T-\theta}{C},
\label{eq:curie_weiss}
\end{equation}
where $\chi_0$ is a temperature-independent term~\cite{ref27_erickson2007}, $C$ is the Curie constant, and $\theta$ is the Weiss temperature.
The fitting yields $\theta$ = 9.8~K at $x$=0.04. Field-dependent magnetization $M(H)$ for these samples is provided in Fig.~S1 (Supplementary Material).
The saturation moment $M_{\mathrm{sat}}$ follows the same trend as $T_{\mathrm{C}}$, reaching a maximum of $\sim$0.93~$\mu_{\mathrm{B}}$/Mn at $x$=0.04 (inset Fig. 2a). The Curie temperature $T_{\mathrm{C}}$ discussed below was determined independently from the onset of the magnetization anomaly, as indicated in Fig. 2a. Further increasing K suppresses both $T_{\mathrm{C}}$ and $M_{\mathrm{sat}}$.

This ``optimal carrier density'' behavior is consistent with the broader view that ferromagnetism in many DMS systems is mediated by carriers (often discussed in RKKY/Zener-type frameworks)~\cite{ref20_glasbrenner2014,ref23_dietl2000science}.
In the present material, K substitution introduces hole carriers and thereby tunes ferromagnetic ordering and the effective Mn--Mn interactions, similar in spirit to other decoupled charge/spin-doped DMS platforms~\cite{ref9_deng2011,ref10_zhao2013,ref11_zhao2014csb,ref14_chen2014prb,ref30_deng2013}.

\emph{Mn-doping dependence: soft ferromagnetism and competing interactions}

Figure~2b shows magnetic hysteresis loops at 2~K for (Ba$_{0.96}$K$_{0.04}$)(Cd$_{1-y}$Mn$_y$)$_2$As$_2$ with $y$=0.05, 0.07, 0.10, and 0.20. Small coercive fields ($H_{\mathrm{C}}$) are observed, which are favorable for low-field magnetic switching.
With increasing Mn content, $M_{\mathrm{sat}}$ decreases from $0.94\,\mu_{\mathrm{B}}/\mathrm{Mn}$ to $0.34\,\mu_{\mathrm{B}}/\mathrm{Mn}$, while $H_{\mathrm{C}}$ increases modestly from $10\,\mathrm{Oe}$ to $30\,\mathrm{Oe}$. These values are much smaller than the $5\,\mu_{\mathrm{B}}/\mathrm{Mn}$ expected for fully aligned Mn$^{2+}$ local moments ($S = 5/2$), indicating that only a fraction of the Mn moments contributes to the net ferromagnetic order. The reduction becomes more pronounced at higher Mn content, consistent with an increasing probability of antiferromagnetically coupled Mn--Mn near-neighbor pairs, which compete with the carrier-mediated ferromagnetic interaction.
The inset of Fig.~2b shows that the onset-defined $T_{\mathrm{C}}$ reaches a maximum of $\sim 17$~K at $y = 0.20$ (Fig.~S4). For heavier Mn substitution, $T_{\mathrm{C}}$ gradually decreases, as summarized in Fig.~2c.
Similar non-monotonic $T_{\mathrm{C}}(y)$ behavior has been reported in multiple Mn-doped DMS systems~\cite{ref3_ohno1998,ref9_deng2011,ref11_zhao2014csb,ref14_chen2014prb,ref29_chen2016scirep,ref30_deng2013,ref31_chen2016aipadv}.

A plausible interpretation is competition between (i) carrier-regulated ferromagnetic interactions among randomly distributed Mn moments and (ii) increasing antiferromagnetic superexchange associated with Mn--Mn near-neighbor pairs at higher Mn concentrations, in line with theoretical expectations for Mn-doped II--II--V and related DMS chemistries~\cite{ref20_glasbrenner2014}.
The relevance of Mn--Mn antiferromagnetic coupling is also underscored by the strong antiferromagnetism of Mn-based parent compounds in related structural motifs (e.g., BaMn$_2$As$_2$-derived systems)~\cite{ref28_pandey2012}.

\emph{Hall effect: p-type carriers and anomalous Hall response}

An anomalous Hall effect (AHE) is widely used as evidence that ferromagnetism couples to itinerant carriers in DMS systems~\cite{ref8_jungwirth2006,ref23_dietl2000science}.
Here, Hall measurements provide both carrier information and an intrinsic transport signature.
For the nominal $x = 0.04$ sample, the normal Hall response indicates a hole concentration of order $2.8 \times 10^{19}\ \mathrm{cm^{-3}}$. This value is about one order of magnitude smaller than the nominal hole density expected from full K substitution ($\sim 3 \times 10^{20}\ \mathrm{cm^{-3}}$, assuming one hole per K), suggesting that the itinerant carrier density is reduced relative to the nominal chemical doping, likely due to incomplete activation, localization, and/or compensation. For comparison, in the isostructural CaAl$_2$Si$_2$-type system $(\mathrm{Ca,Na})(\mathrm{Zn,Mn})_2\mathrm{As}_2$~\cite{ref15_zhao2014jap}, $10\%$ Na doping yields $n_p \sim 10^{20}\ \mathrm{cm^{-3}}$; scaling by the doping ratio ($4\%$ vs.\ $10\%$), our measured carrier density is broadly consistent with this trend. Nevertheless, the $p$-type Hall response together with the systematic reduction of resistivity with increasing K content (Fig.~4b) supports K as the primary charge-doping channel. Above $T_{\mathrm{C}}$, the Hall resistivity is linear in magnetic field.

Figure~2d presents the Hall resistivity of (Ba$_{0.96}$K$_{0.04}$)(Cd$_{0.9}$Mn$_{0.1}$)$_2$As$_2$ at 2~K. An AHE component is observed at low fields (arrow) and is highlighted in the top-left inset of Fig.~2d.
The extracted hole concentration decreases monotonically upon cooling (bottom-right inset of Fig.~2d), consistent with semiconducting/activated transport and likely contributing to the strong increase of resistivity at low temperature discussed below.

\emph{Specific heat: thermodynamic signature near $T_{\mathrm{C}}$}

To further probe the ferromagnetic transition, we measured the zero-field specific heat of (Ba$_{0.96}$K$_{0.04}$)(Cd$_{0.9}$Mn$_{0.1}$)$_2$As$_2$ (Fig.~3a). A clear anomaly appears near $T_{\mathrm{C}}$ (arrow), consistent with a paramagnetic--ferromagnetic transition.
The low-temperature specific heat data above $T_{\mathrm{C}}$ are fitted in the range $10\,\mathrm{K} < T < 16\,\mathrm{K}$ using
\begin{equation}
C_{\mathrm{electron}} + C_{\mathrm{lattice}} = \gamma T + \beta_3 T^3 + \beta_5 T^5,
\label{eq:cv_fit}
\end{equation}
yielding $\gamma = 14.94\,\mathrm{mJ\,mol^{-1}\,K^{-2}}$, $\beta_3 = 0.87\,\mathrm{mJ\,mol^{-1}\,K^{-4}}$, and $\beta_5 = 7.94 \times 10^{-4}\,\mathrm{mJ\,mol^{-1}\,K^{-6}}$ (coefficient of determination $R^2 = 0.99991$; inset of Fig.~3a).
After subtracting lattice and electronic contributions, the magnetic contribution $C_M=C_{\mathrm{tot}}-C_{\mathrm{lattice}}-C_{\mathrm{electron}}$ (Fig.~3b) exhibits a broad feature around $T_{\mathrm{C}}$, rather than a sharp $\lambda$-type anomaly expected for an ideal mean-field transition.
Such broadening is plausible in chemically substituted DMS materials where disorder and competing interactions coexist (carrier-regulated ferromagnetism versus Mn--Mn antiferromagnetic tendencies)~\cite{ref20_glasbrenner2014}, and may also reflect a distribution of local magnetic environments typical of substituted magnetic systems~\cite{ref32_bouvier1991}. Field-dependent specific-heat measurements would be valuable for further testing the nature of the magnetic transition through possible field-induced broadening or shifts of the specific-heat feature, although this remains an important direction for future study.

Compared with earlier reports on this compound family that primarily emphasized magnetization and transport~\cite{ref16_yang2013jap}, the thermodynamic signature near $T_{\mathrm{C}}$ provides an additional bulk indicator supporting intrinsic ordering.

\emph{Resistivity: activation behavior and opposite roles of K and Mn doping}

Figure~4a shows the temperature dependence of resistivity $\rho(T)$ for undoped BaCd$_2$As$_2$, confirming semiconducting behavior.
The activation energy $E_a \approx$0.233~eV is obtained by fitting $\rho(T)$ in the range 180--300~K (inset of Fig.~4a) using:
\begin{equation}
\rho(T) = \rho_0 \exp\!\left(\frac{E_a}{k_{\mathrm{B}} T}\right),
\label{eq:rho_activation}
\end{equation}
where $\rho_0$ is a constant and $k_{\mathrm{B}}$ is the Boltzmann constant.
This value is smaller than a previously reported experimental estimate ($E_a$=0.42~eV)~\cite{ref16_yang2013jap}, but close to a theoretical prediction of $\sim$0.3~eV~\cite{ref33_hua2014}.

K substitution substantially reduces resistivity (Fig.~4a), consistent with hole doping.
A similar trend is observed in (Ba$_{1-x}$K$_x$)(Cd$_{0.9}$Mn$_{0.1}$)$_2$As$_2$ (open symbols in Fig.~4b), where increasing $x$ monotonically decreases $\rho(T)$.
In contrast, increasing Mn content in (Ba$_{0.96}$K$_{0.04}$)(Cd$_{1-y}$Mn$_y$)$_2$As$_2$ (solid symbols in Fig.~4b) increases resistivity, plausibly reflecting enhanced disorder/localization and spin-dependent scattering introduced by Mn dopants.

\emph{Magnetoresistance: evolution to colossal $MR$ at high Mn content and performance benchmarking}

Negative magnetoresistance (defined as $MR=[\rho(H)-\rho(0)]/\rho(0)$) below $T_{\mathrm{C}}$ is shown in Fig.~S2a. In the present work, the term ``colossal'' refers to negative $MR$ approaching $-100\%$ under this convention; for $(\mathrm{Ba}_{0.96}\mathrm{K}_{0.04})(\mathrm{Cd}_{0.7}\mathrm{Mn}_{0.3})_2\mathrm{As}_2$, the $MR$ at $2\,\mathrm{K}$ is approximately $-99.5\%$. Field-dependent resistivity $\rho(H)$ for (Ba$_{0.96}$K$_{0.04}$)(Cd$_{0.85}$Mn$_{0.15}$)$_2$As$_2$ at several temperatures is shown in Fig.~S2b.
A small positive $MR$ appears at low fields, which may be related to spin reorientation/rotation processes and has also been discussed for ferromagnetic semiconductors such as (Ga,Mn)As~\cite{ref34_matsukura1998}.

A large negative $MR$ was previously reported for (Ba,K)(Cd,Mn)$_2$As$_2$ with Mn content up to $y$=0.2~\cite{ref16_yang2013jap}.
Here we extend to substantially heavier Mn substitution (0.05$\le$ $y$ $\le$0.5) and observe a pronounced enhancement of $MR$ in the high-$y$ regime.
As shown in Fig.~4c, the magnitude of negative $MR$ increases strongly with Mn content, and colossal negative $MR$ becomes evident at $y$=0.3, approaching $-100\%$ at low temperature.

Figure~4d shows $\rho(T)$ of (Ba$_{0.96}$K$_{0.04}$)(Cd$_{0.7}$Mn$_{0.3}$)$_2$As$_2$ measured under various magnetic fields.
The $MR$ grows rapidly upon cooling, and the resistivity becomes very large at low temperature, consistent with an insulating/localized transport regime in which magnetic field can strongly modify carrier scattering and/or localization.
At 2~K, the negative $MR$ approaches $-100\%$ and nearly saturates at a relatively low magnetic field of $\sim$0.35~T (inset of Fig.~4d).
This low-field saturation is attractive from a performance perspective because it suggests that large resistivity modulation can be achieved without requiring multi-tesla fields.

The strong low-temperature enhancement of negative $MR$ at high $y$ is consistent with field-suppressed spin disorder and/or field-assisted delocalization scenarios that have been discussed for magnetic semiconductors and related disordered magnetic conductors~\cite{ref35_kobayashi1998,ref36_vonmolnar1983}.
Because the present samples are polycrystalline, microstructure (e.g., grain boundaries) may further amplify the field response. For the $y = 0.3$ composition, the magnetization becomes nearly saturated within a few hundred Oe, whereas the $MR$ continues to evolve strongly up to $\sim 3500$~Oe. This large mismatch in characteristic field scales suggests that spin-disorder scattering alone does not fully account for the colossal negative $MR$. Additional mechanisms, such as field-assisted carrier delocalization and grain-boundary effects in the polycrystalline samples, likely contribute in the heavily Mn-doped, insulating regime.
Importantly, the largest $MR$ is observed in compositions where the net moment per Mn is most strongly reduced and the resistivity is highest. This correlation is physically self-consistent: heavy Mn substitution enhances disorder/localization and strengthens competing magnetic interactions, both of which increase the sensitivity of transport to an applied magnetic field. As a result, the Mn-rich regime exhibits the strongest magnetoresistive response even though the net ordered moment per Mn is reduced. More broadly, increasing Mn enhances disorder/localization and strengthens the $MR$ response, while K doping reduces the baseline resistivity by increasing hole density.

Finally, both (Ba$_{0.96}$K$_{0.04}$)(Cd$_{0.6}$Mn$_{0.4}$)$_2$As$_2$ and (Ba$_{0.96}$K$_{0.04}$)(Cd$_{0.5}$Mn$_{0.5}$)$_2$As$_2$ exhibit similar behavior (Figs.~S2c and S2d), supporting the reproducibility of the colossal $MR$ trend in the heavily Mn-doped window.
Colossal negative $MR$ has also been reported in other CaAl$_2$Si$_2$-type bulk DMS systems such as (Sr,K)(Zn,Mn)$_2$As$_2$~\cite{ref18_yang2014epl}.
The comparable $MR$ magnitude, together with low-field saturation in (Ba,K)(Cd,Mn)$_2$As$_2$, strengthens the case that layered, decoupled-doping bulk DMS materials can serve as chemically tunable platforms for magnetoresistive functionalities at low temperature.

\subsection*{Conclusions}

In summary, we synthesized and systematically investigated the structural, magnetic, thermodynamic, and magnetotransport properties of the layered CaAl$_2$Si$_2$-type diluted magnetic semiconductor (Ba$_{1-x}$K$_x$)(Cd$_{1-y}$Mn$_y$)$_2$As$_2$ over a broad Mn substitution range of 0.05$\le$ $y$ $\le$0.5.
Powder XRD confirms that all compositions retain the same hexagonal structure, and the monotonic evolution of lattice parameters with K and Mn substitution supports successful chemical doping.

Ferromagnetic ordering is realized through decoupled charge (K) and spin (Mn) doping.
The Curie temperature and saturation moment are optimized at intermediate carrier doping: $T_{\mathrm{C}}$ reaches 9.8~K at $x$=0.04 for fixed $y$=0.10, while the Mn-doping series at fixed $x$=0.04 shows a maximum $T_C \approx$17~K at $y$=0.20, followed by a gradual decrease at higher Mn contents.
The intrinsic nature of magnetic ordering is further supported by an anomalous Hall contribution at low fields and a specific-heat anomaly near $T_{\mathrm{C}}$.

Most importantly from a materials-performance perspective, the magnetoresistance is strongly enhanced in the heavily Mn-doped regime.
For $y \ge$0.3, the system exhibits colossal negative magnetoresistance approaching $-100\%$ at 2~K, with near-saturation at a relatively low magnetic field of $\sim$0.35~T.
This low-field saturation, together with soft ferromagnetism and clear intrinsic magnetic signatures, highlights (Ba,K)(Cd,Mn)$_2$As$_2$ as a chemically tunable bulk DMS platform for low-temperature magnetoresistive functionalities, and provides guidance for optimizing $MR$ through composition-controlled carrier density and magnetic disorder.

\vspace{3mm}

\emph{Acknowledgments}---This work was supported by the National Science Foundation and the Ministry of Science and Technology (MOST) of China.

\vspace{3mm}

\emph{Author Contributions}---C.~Q.~J. and B.~J.~C. conceived and coordinated the work. B.~J.~C. designed and grew the polycrystalline (Ba,K)(Cd,Mn)$_2$As$_2$ samples, performed preliminary characterizations, and carried out the magnetic susceptibility and electronic transport measurements with the help of Z.~D. B.~J.~C. analyzed the data. C.~Q.~J., B.~J.~C., and Z.~D. wrote the manuscript.

\vspace{3mm}

\emph{Competing interests}---The authors declare no competing interests.

\vspace{3mm}

\emph{Data availability}---The data that support the findings of this study are available from the corresponding author upon reasonable request.

\bibliographystyle{elsarticle-num}
\bibliography{sn-bibliography}

\begin{thebibliography}{10}
\expandafter\ifx\csname url\endcsname\relax
  \def\url#1{\texttt{#1}}\fi
\expandafter\ifx\csname urlprefix\endcsname\relax\def\urlprefix{URL }\fi
\expandafter\ifx\csname href\endcsname\relax
  \def\href#1#2{#2} \def\path#1{#1}\fi

\bibitem{ref1_awschalom2007}
D.~D. Awschalom, M.~E. Flatt{\'e}, {Challenges for semiconductor spintronics}, Nature Physics 3~(3) (2007) 153--159.
\newblock \href {https://doi.org/10.1038/nphys551} {\path{doi:10.1038/nphys551}}.

\bibitem{ref2_zutic2004}
I.~{\v{Z}}uti{\'c}, J.~Fabian, S.~Das~Sarma, {Spintronics: Fundamentals and applications}, Reviews of Modern Physics 76~(2) (2004) 323--410.
\newblock \href {https://doi.org/10.1103/RevModPhys.76.323} {\path{doi:10.1103/RevModPhys.76.323}}.

\bibitem{ref4_dietl2010}
T.~Dietl, {A ten-year perspective on dilute magnetic semiconductors and oxides}, Nature Materials 9~(12) (2010) 965--974.
\newblock \href {https://doi.org/10.1038/nmat2898} {\path{doi:10.1038/nmat2898}}.

\bibitem{ref3_ohno1998}
H.~Ohno, {Making nonmagnetic semiconductors ferromagnetic}, Science 281~(5379) (1998) 951--956.
\newblock \href {https://doi.org/10.1126/science.281.5379.951} {\path{doi:10.1126/science.281.5379.951}}.

\bibitem{ref5_ohno1996}
H.~Ohno, A.~Shen, F.~Matsukura, A.~Oiwa, A.~Endo, S.~Katsumoto, Y.~Iye, {(Ga,Mn)As: A new diluted magnetic semiconductor based on GaAs}, Applied Physics Letters 69~(3) (1996) 363--365.
\newblock \href {https://doi.org/10.1063/1.118061} {\path{doi:10.1063/1.118061}}.

\bibitem{ref8_jungwirth2006}
T.~Jungwirth, J.~Sinova, J.~Ma{\v{s}}ek, J.~Ku{\v{c}}era, A.~H. MacDonald, {Theory of ferromagnetic (III,Mn)V semiconductors}, Reviews of Modern Physics 78~(3) (2006) 809--864.
\newblock \href {https://doi.org/10.1103/RevModPhys.78.809} {\path{doi:10.1103/RevModPhys.78.809}}.

\bibitem{ref34_matsukura1998}
F.~Matsukura, H.~Ohno, A.~Shen, Y.~Sugawara, {Transport properties and origin of ferromagnetism in (Ga,Mn)As}, Physical Review B 57 (1998) R2037--R2040.
\newblock \href {https://doi.org/10.1103/PhysRevB.57.R2037} {\path{doi:10.1103/PhysRevB.57.R2037}}.

\bibitem{ref13_wang2008}
M.~Wang, R.~P. Campion, A.~W. Rushforth, K.~W. Edmonds, C.~T. Foxon, B.~L. Gallagher, {Achieving high {Curie} temperature in (Ga,Mn)As}, Applied Physics Letters 93~(13) (2008) 132103.
\newblock \href {https://doi.org/10.1063/1.2992200} {\path{doi:10.1063/1.2992200}}.

\bibitem{ref12_chen2011}
L.~Chen, X.~Yang, F.~Yang, J.~Zhao, J.~Misuraca, P.~Xiong, S.~von Moln{\'a}r, {Enhancing the {Curie} temperature of ferromagnetic semiconductor (Ga,Mn)As to 200 {K} via nanostructure engineering}, Nano Letters 11~(7) (2011) 2584--2589.
\newblock \href {https://doi.org/10.1021/nl201187m} {\path{doi:10.1021/nl201187m}}.

\bibitem{ref6_samarth2010}
N.~Samarth, {A model ferromagnetic semiconductor}, Nature Materials 9~(12) (2010) 955--956.
\newblock \href {https://doi.org/10.1038/nmat2908} {\path{doi:10.1038/nmat2908}}.

\bibitem{ref7_chambers2010}
S.~Chambers, {Is it really intrinsic ferromagnetism?}, Nature Materials 9~(12) (2010) 956--957.
\newblock \href {https://doi.org/10.1038/nmat2905} {\path{doi:10.1038/nmat2905}}.

\bibitem{ref9_deng2011}
Z.~Deng, C.~Q. Jin, Q.~Q. Liu, et~al., {Li(Zn,Mn)As as a new generation ferromagnet based on a I--II--V semiconductor}, Nature Communications 2 (2011) 422.
\newblock \href {https://doi.org/10.1038/ncomms1425} {\path{doi:10.1038/ncomms1425}}.

\bibitem{ref30_deng2013}
Z.~Deng, et~al., {Diluted ferromagnetic semiconductor Li(Zn,Mn)P with decoupled charge and spin doping}, Physical Review B 88 (2013) 081203.
\newblock \href {https://doi.org/10.1103/PhysRevB.88.081203} {\path{doi:10.1103/PhysRevB.88.081203}}.

\bibitem{ref10_zhao2013}
K.~Zhao, Z.~Deng, X.~C. Wang, et~al., {New diluted ferromagnetic semiconductor with {Curie} temperature up to 180 {K} and isostructural to the ``122'' iron-based superconductors}, Nature Communications 4 (2013) 1442.
\newblock \href {https://doi.org/10.1038/ncomms2447} {\path{doi:10.1038/ncomms2447}}.

\bibitem{ref11_zhao2014csb}
K.~Zhao, B.~Chen, G.~Zhao, Z.~Yuan, Q.~Liu, Z.~Deng, J.~Zhu, C.~Jin, {Ferromagnetism at 230 K in (Ba$_{0.7}$K$_{0.3}$)(Zn$_{0.85}$Mn$_{0.15}$)$_2$As$_2$ diluted magnetic semiconductor}, Chinese science bulletin 59~(21) (2014) 2524--2527.
\newblock \href {https://doi.org/10.1007/s11434-014-0398-z} {\path{doi:10.1007/s11434-014-0398-z}}.

\bibitem{ref19_man2015}
H.~Man, S.~Guo, Y.~Sui, Y.~Guo, B.~Chen, H.~Wang, C.~Ding, F.~L. Ning, {Ba(Zn$_{1-2x}$Mn$_x$Cu$_x$)$_2$As$_2$: A bulk form diluted ferromagnetic semiconductor with Mn and Cu codoping at Zn sites}, Scientific Reports 5 (2015) 15507.
\newblock \href {https://doi.org/10.1038/srep15507} {\path{doi:10.1038/srep15507}}.

\bibitem{ref21_suzuki2015prb}
H.~Suzuki, K.~Zhao, G.~Shibata, Y.~Takahashi, S.~Sakamoto, K.~Yoshimatsu, B.~J. Chen, H.~Kumigashira, F.~H. Chang, H.~J. Lin, et~al., {Photoemission and x-ray absorption studies of the diluted magnetic semiconductor Ba$_{1-x}$K$_x$(Zn$_{1-y}$Mn$_y$)$_2$As$_2$ isostructural to {Fe}-based superconductors}, Physical Review B 91 (2015) 140401.
\newblock \href {https://doi.org/10.1103/PhysRevB.91.140401} {\path{doi:10.1103/PhysRevB.91.140401}}.

\bibitem{ref24_suzuki2015prb_arpes}
H.~Suzuki, et~al., {Fermi surfaces and $p$--$d$ hybridization in the diluted magnetic semiconductor Ba$_{1-x}$K$_x$(Zn$_{1-y}$Mn$_y$)$_2$As$_2$ studied by soft x-ray angle-resolved photoemission spectroscopy}, Physical Review B 92 (2015) 235120.
\newblock \href {https://doi.org/10.1103/PhysRevB.92.235120} {\path{doi:10.1103/PhysRevB.92.235120}}.

\bibitem{ref14_chen2014prb}
B.~J. Chen, et~al., {(Sr,Na)(Zn,Mn)$_2$As$_2$: a diluted ferromagnetic semiconductor with the hexagonal {CaAl}$_2${Si}$_2$ type structure}, Physical Review B 90 (2014) 155202.
\newblock \href {https://doi.org/10.1103/PhysRevB.90.155202} {\path{doi:10.1103/PhysRevB.90.155202}}.

\bibitem{ref15_zhao2014jap}
K.~Zhao, B.~J. Chen, Z.~Deng, et~al., {(Ca,Na)(Zn,Mn)$_2$As$_2$: a new spin and charge doping decoupled diluted ferromagnetic semiconductor}, Journal of Applied Physics 116 (2014) 163906.
\newblock \href {https://doi.org/10.1063/1.4899190} {\path{doi:10.1063/1.4899190}}.

\bibitem{ref17_chen2016jap}
B.~Chen, Z.~Deng, W.~Li, M.~Gao, Z.~Li, G.~Zhao, S.~Yu, X.~Wang, Q.~Liu, C.~Jin, {(Sr$_{1-x}$Na$_x$)(Cd$_{1-x}$Mn$_x$)$_2$As$_2$: A new charge and spin doping decoupled diluted magnetic semiconductors with {CaAl}$_2${Si}$_2$-type structure}, Journal of Applied Physics 120~(8) (2016) 083902.
\newblock \href {https://doi.org/10.1063/1.4961565} {\path{doi:10.1063/1.4961565}}.

\bibitem{ref18_yang2014epl}
X.~Yang, Q.~Chen, Y.~Li, Z.~Wang, J.~Bao, Y.~Li, Q.~Tao, G.~Cao, Z.-A. Xu, {Sr$_{0.9}$K$_{0.1}$Zn$_{1.8}$Mn$_{0.2}$As$_2$: A ferromagnetic semiconductor with colossal magnetoresistance}, Europhysics Letters (EPL) 107 (2014) 67007.
\newblock \href {https://doi.org/10.1209/0295-5075/107/67007} {\path{doi:10.1209/0295-5075/107/67007}}.

\bibitem{ref22_yang2016jmmm}
J.~Yang, S.~Luo, Y.~Xiong, {Magnetic mechanism investigations on K and Mn co-doped diluted magnetic semiconductor (Sr,K)(Zn,Mn)$_2$As$_2$}, Journal of Magnetism and Magnetic Materials 407 (2016) 334--340.
\newblock \href {https://doi.org/10.1016/j.jmmm.2016.02.012} {\path{doi:10.1016/j.jmmm.2016.02.012}}.

\bibitem{ref40_yu2019aplmater}
S.~Yu, G.~Zhao, Y.~Peng, X.~Zhu, X.~Wang, J.~Zhao, L.~Cao, W.~Li, Z.~Li, Z.~Deng, C.~Jin, {A substantial increase of {Curie} temperature in a new type of diluted magnetic semiconductors via effects of chemical pressure}, APL Materials 7~(10) (2019) 101119.
\newblock \href {https://doi.org/10.1063/1.5120719} {\path{doi:10.1063/1.5120719}}.

\bibitem{ref41_dong2022jsemicond}
J.~Dong, X.~Zhao, L.~Fu, Y.~Gu, R.~Zhang, Q.~Yang, L.~Xie, F.~Ning, {(Ca,K)(Zn,Mn)$_2$As$_2$: Ferromagnetic semiconductor induced by decoupled charge and spin doping in CaZn$_2$As$_2$}, Journal of Semiconductors 43~(7) (2022) 072501.
\newblock \href {https://doi.org/10.1088/1674-4926/43/7/072501} {\path{doi:10.1088/1674-4926/43/7/072501}}.

\bibitem{ref42_yu2023jsemicond}
S.~Yu, Y.~Peng, G.~Zhao, J.~Zhao, X.~Wang, J.~Zhang, Z.~Deng, C.~Jin, {Colossal negative magnetoresistance in spin glass Na(Zn,Mn)Sb}, Journal of Semiconductors 44~(3) (2023) 032501.
\newblock \href {https://doi.org/10.1088/1674-4926/44/3/032501} {\path{doi:10.1088/1674-4926/44/3/032501}}.

\bibitem{ref44_zhao2025prb}
G.~Zhao, X.~Li, S.~Yu, H.~Chen, Y.~Hu, Y.~Cai, S.~Guo, B.~Gu, F.~Ning, K.~M. Kojima, Z.~Deng, Y.~Xing, H.-J. Gao, X.~Zhou, G.~Su, S.~Maekawa, C.~Jin, Y.~J. Uemura, {Spin-charge interplay in the diluted magnetic semiconductor Na(Zn,Mn)Sb studied by multiprobe measurements and simulations}, Physical Review B 112 (2025) 075104.
\newblock \href {https://doi.org/10.1103/PhysRevB.112.075104} {\path{doi:10.1103/PhysRevB.112.075104}}.

\bibitem{ref43_peng2024nanomaterials}
Y.~Peng, L.~Shi, G.~Zhao, J.~Zhang, J.~Zhao, X.~Wang, Z.~Deng, C.~Jin, {Colossal magnetoresistance in layered diluted magnetic semiconductor Rb(Zn,Li,Mn)$_4$As$_3$ single crystals}, Nanomaterials 14~(3) (2024) 263.
\newblock \href {https://doi.org/10.3390/nano14030263} {\path{doi:10.3390/nano14030263}}.

\bibitem{ref16_yang2013jap}
X.~Yang, Y.~Li, P.~Zhang, H.~Jiang, Y.~Luo, Q.~Chen, C.~Feng, C.~Cao, J.~Dai, Q.~Tao, G.~Cao, Z.~Xu, {K and Mn co-doped BaCd$_2$As$_2$: a hexagonal structured bulk diluted magnetic semiconductor with large magnetoresistance}, Journal of Applied Physics 114 (2013) 223905.
\newblock \href {https://doi.org/10.1063/1.4842875} {\path{doi:10.1063/1.4842875}}.

\bibitem{ref37_chen2016cpb}
B.~Chen, Z.~Deng, X.~Wang, S.~Feng, Z.~Yuan, S.~Zhang, Q.~Liu, C.~Jin, {Structural stability at high pressure, electronic, and magnetic properties of BaFZnAs: A new candidate of host material of diluted magnetic semiconductors}, Chinese Physics B 25~(7) (2016) 077503.
\newblock \href {https://doi.org/10.1088/1674-1056/25/7/077503} {\path{doi:10.1088/1674-1056/25/7/077503}}.

\bibitem{ref38_larson1994gsas}
A.~C. Larson, R.~B. Von~Dreele, {{GSAS}: {General Structure Analysis System}}, Los Alamos National Laboratory Report LAUR 86-748, los Alamos, New Mexico (1994).

\bibitem{ref25_klufers1984}
P.~Kl{\"u}fers, H.~Neumann, A.~Mewis, H.-U. Schuster, {AB2X2-verbindungen im CaAl2Si2-Typ, VIII [1]/AB2X2 compounds with the CaAl2Si2 structure, VIII [1]}, Zeitschrift f{\"u}r Naturforschung B 35~(10) (1980) 1317--1318.

\bibitem{ref26_mewis1980}
A.~Mewis, {Ternary phosphides with the {ThCr}$_2${Si}$_2$ structure}, Zeitschrift f{\"u}r Naturforschung B 35 (1980) 141.
\newblock \href {https://doi.org/10.1515/znb-1980-0205} {\path{doi:10.1515/znb-1980-0205}}.

\bibitem{ref27_erickson2007}
A.~Erickson, et~al., {Ferromagnetism in the {Mott} insulator Ba$_2$NaOsO$_6$}, Physical Review Letters 99 (2007) 016404.
\newblock \href {https://doi.org/10.1103/PhysRevLett.99.016404} {\path{doi:10.1103/PhysRevLett.99.016404}}.

\bibitem{ref20_glasbrenner2014}
J.~K. Glasbrenner, I.~{\v{Z}}uti{\'c}, I.~I. Mazin, {Theory of Mn-doped {II-II-V} semiconductors}, Physical Review B 90 (2014) 140403.
\newblock \href {https://doi.org/10.1103/PhysRevB.90.140403} {\path{doi:10.1103/PhysRevB.90.140403}}.

\bibitem{ref23_dietl2000science}
T.~Dietl, H.~Ohno, F.~Matsukura, J.~Cibert, D.~Ferrand, {Zener model description of ferromagnetism in zinc-blende magnetic semiconductors}, Science 287~(5455) (2000) 1019--1022.
\newblock \href {https://doi.org/10.1126/science.287.5455.1019} {\path{doi:10.1126/science.287.5455.1019}}.

\bibitem{ref29_chen2016scirep}
B.~J. Chen, et~al., {New fluoride-arsenide diluted magnetic semiconductor (Ba,K)F(Zn,Mn)As with independent spin and charge doping}, Scientific Reports 6 (2016) 36578.
\newblock \href {https://doi.org/10.1038/srep36578} {\path{doi:10.1038/srep36578}}.

\bibitem{ref31_chen2016aipadv}
B.~Chen, Z.~Deng, W.~Li, M.~Gao, J.~Zhao, G.~Zhao, S.~Yu, X.~Wang, Q.~Liu, C.~Jin, {Li(Zn,Co,Mn)As: A bulk form diluted magnetic semiconductor with Co and Mn co-doping at Zn sites}, AIP Advances 6~(11) (2016) 115014.
\newblock \href {https://doi.org/10.1063/1.4967778} {\path{doi:10.1063/1.4967778}}.

\bibitem{ref28_pandey2012}
A.~Pandey, et~al., {Ba$_{1-x}$K$_x$Mn$_2$As$_2$: An antiferromagnetic local-moment metal}, Physical Review Letters 108 (2012) 087005.
\newblock \href {https://doi.org/10.1103/PhysRevLett.108.087005} {\path{doi:10.1103/PhysRevLett.108.087005}}.

\bibitem{ref32_bouvier1991}
M.~Bouvier, P.~Lethuillier, D.~Schmitt, {Specific heat in some gadolinium compounds. I. Experimental}, Physical Review B 43 (1991) 13137.
\newblock \href {https://doi.org/10.1103/PhysRevB.43.13137} {\path{doi:10.1103/PhysRevB.43.13137}}.

\bibitem{ref33_hua2014}
L.~Hua, Q.~L. Zhu, {Electronic structure and magnetic properties of diluted magnetic semiconductor K and Mn co-doped BaCd$_2$As$_2$ from first-principles calculations}, Modern Physics Letters B 28 (2014) 1450111.
\newblock \href {https://doi.org/10.1142/S0217984914501115} {\path{doi:10.1142/S0217984914501115}}.

\bibitem{ref35_kobayashi1998}
K.-I. Kobayashi, T.~Kimura, H.~Sawada, K.~Terakura, Y.~Tokura, {Room-temperature magnetoresistance in an oxide material with an ordered double-perovskite structure}, Nature 395~(6703) (1998) 677--680.
\newblock \href {https://doi.org/10.1038/27167} {\path{doi:10.1038/27167}}.

\bibitem{ref36_vonmolnar1983}
S.~von Moln{\'a}r, A.~Briggs, J.~Flouquet, G.~Remenyi, {Electron localization in a magnetic semiconductor: Gd$_{3-x}v_x$S$_4$}, Physical Review Letters 51~(8) (1983) 706--709.
\newblock \href {https://doi.org/10.1103/PhysRevLett.51.706} {\path{doi:10.1103/PhysRevLett.51.706}}.

\end{thebibliography}

\newpage

\begin{figure}
    \centering
    \includegraphics[width=1.0\linewidth]{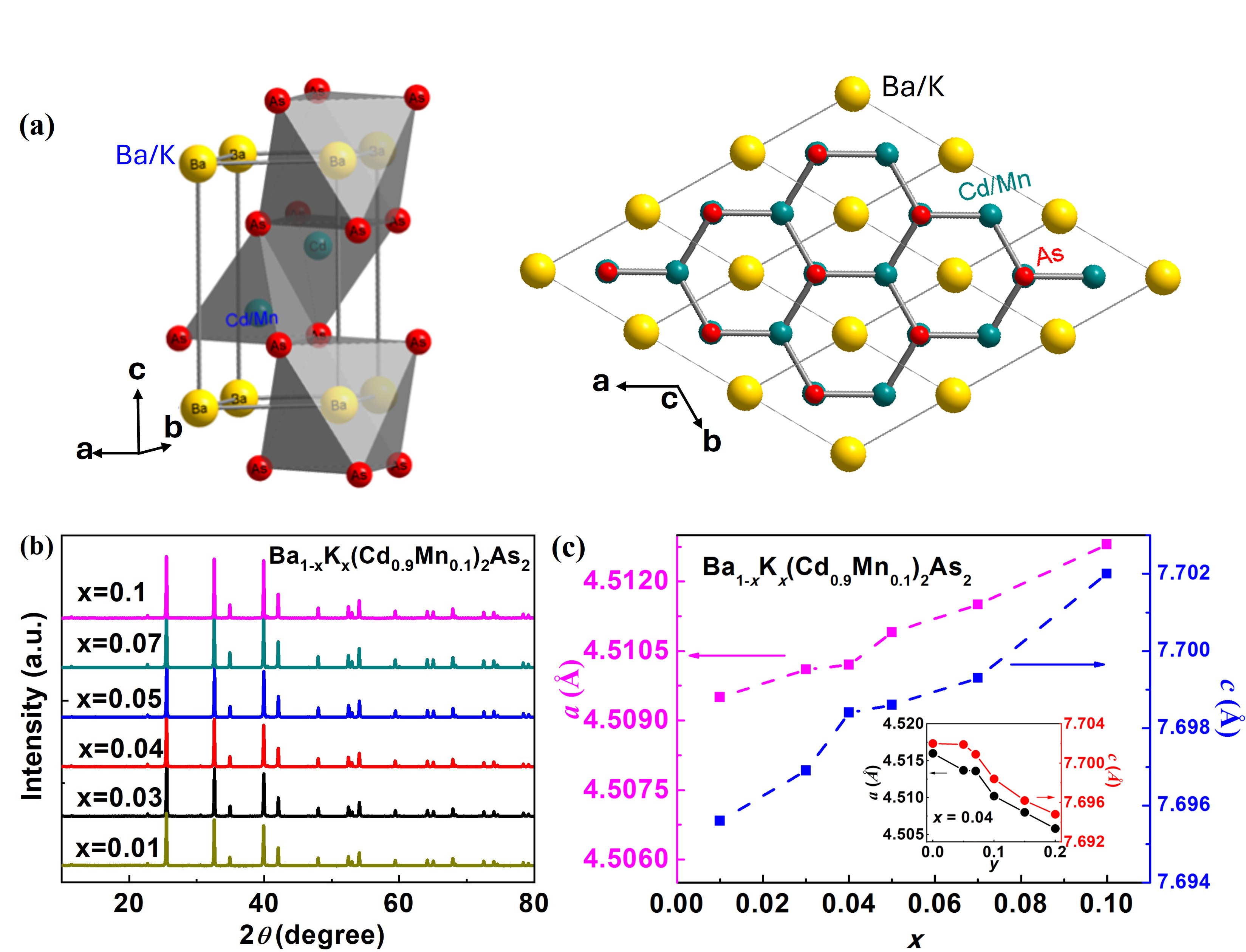}
    \caption{
    \textbf{Crystal structure and phase characterization of CaAl$_2$Si$_2$-type (Ba,K)(Cd,Mn)$_2$As$_2$.}
    \textbf{a}
    Crystal structure of (Ba,K)(Cd,Mn)$_2$As$_2$ (space group $P\bar{3}m1$), composed of CdAs$_4$ tetrahedra and BaAs$_6$ octahedra and forming a layered Cd$_2$As$_2$ network.
    \textbf{b}
    Powder X-ray diffraction patterns of (Ba$_{1-x}$K$_x$)(Cd$_{0.9}$Mn$_{0.1}$)$_2$As$_2$ for $x$=0.01, 0.03, 0.04, 0.05, 0.07, and 0.10. All patterns can be indexed by the CaAl$_2$Si$_2$-type structure, indicating no structural transition across the measured K-doping range.
    \textbf{c}
    Lattice parameters refined from XRD as a function of composition. Both $a$ and $c$ expand with K substitution, whereas Mn substitution leads to a contraction (inset), supporting successful chemical doping.}
    \label{fig:fig1}
\end{figure}

\newpage
\begin{figure}
    \centering
    \includegraphics[width=1.0\linewidth]{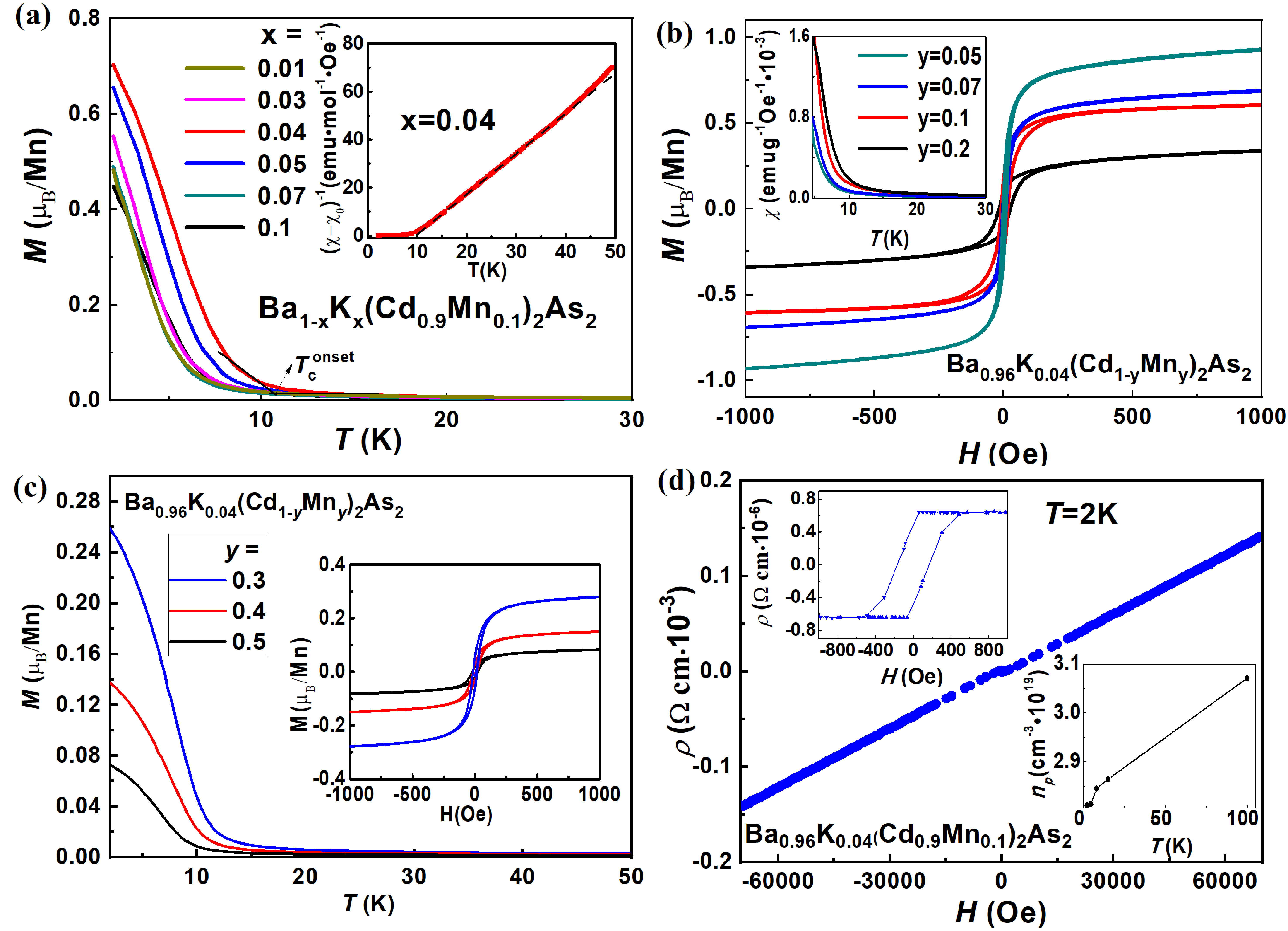}
    \caption{
    \textbf{Ferromagnetism and Hall response in (Ba,K)(Cd,Mn)$_2$As$_2$ with decoupled charge and spin doping.}
    \textbf{a}
    Temperature dependence of magnetization $M(T)$ measured under 500~Oe for the K-doping series (Ba$_{1-x}$K$_x$)(Cd$_{0.9}$Mn$_{0.1}$)$_2$As$_2$ with $x$=0.01--0.10, showing ferromagnetic ordering; arrows indicate the onset-defined $T_{\mathrm{C}}$. Inset: Curie--Weiss analysis in the paramagnetic regime.
    \textbf{b}
    Magnetic hysteresis loops $M(H)$ at 2~K for the Mn-doping series (Ba$_{0.96}$K$_{0.04}$)(Cd$_{1-y}$Mn$_y$)$_2$As$_2$ with $y$=0.05, 0.07, 0.10, and 0.20, demonstrating soft ferromagnetism with small coercive fields; inset: $T_{\mathrm{C}}$ evolution with Mn content.
    \textbf{c}
    Temperature dependence of magnetization for heavily Mn-doped compositions (representative high-$y$ samples), illustrating the reduction of $T_{\mathrm{C}}$ at higher Mn concentrations.
    \textbf{d}
    Hall resistivity $\rho_{xy}(H)$ of (Ba$_{0.96}$K$_{0.04}$)(Cd$_{0.9}$Mn$_{0.1}$)$_2$As$_2$ at 2~K, showing an anomalous Hall contribution at low fields (top-left inset); bottom-right inset: temperature dependence of hole concentration extracted from the normal Hall coefficient, indicating $p$-type carriers with $n_p \sim 2.8 \times 10^{19}$~cm$^{-3}$.}
    \label{fig:fig2}
\end{figure}

\newpage
\begin{figure}
    \centering
    \includegraphics[width=1.0\linewidth]{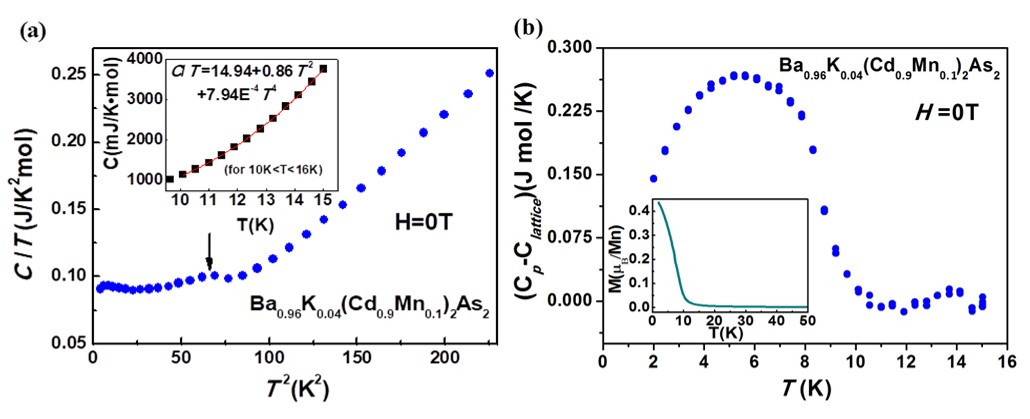}
    \caption{
    \textbf{Specific-heat evidence supporting magnetic ordering.}
    \textbf{a}
    Zero-field specific heat plotted as $C/T$ versus $T^2$ (note: the $x$-axis is $T^2$, not $T$) for $(\mathrm{Ba}_{0.96}\mathrm{K}_{0.04})(\mathrm{Cd}_{0.9}\mathrm{Mn}_{0.1})_2\mathrm{As}_2$, showing an anomaly near $T_{\mathrm{C}}$ (arrow, corresponding to $T^2 \approx 70$--$96~\mathrm{K}^2$, i.e., $T \approx 8.4$--$9.8~\mathrm{K}$). Inset: low-temperature fit using $C_{\mathrm{electron}} + C_{\mathrm{lattice}} = \gamma T + \beta_3 T^3 + \beta_5 T^5$, yielding $\gamma = 14.94~\mathrm{mJ\,mol^{-1}\,K^{-2}}$. 
    \textbf{b} 
    Magnetic contribution $C_M(T)$ obtained after subtracting electronic and lattice terms, plotted versus $T$, exhibiting a broad feature around $T_{\mathrm{C}} \approx 9.8~\mathrm{K}$; inset: susceptibility/magnetization data for the same composition, consistent with the transition temperature.}
    \label{fig:fig3}
\end{figure}

\newpage
\begin{figure}
    \centering
    \includegraphics[width=1.0\linewidth]{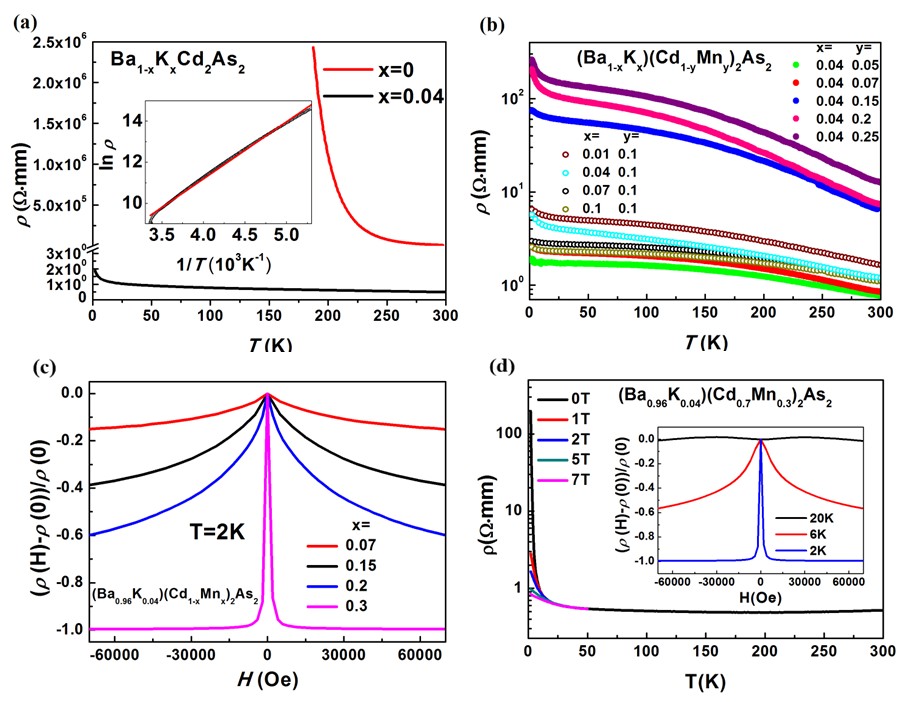}
    \caption{
    \textbf{Transport evolution and colossal low-field negative magnetoresistance in (Ba,K)(Cd,Mn)$_2$As$_2$.}
    \textbf{a}
    Temperature dependence of resistivity $\rho(T)$ for the parent compound BaCd$_2$As$_2$ and representative K-doped sample(s); inset: Arrhenius plot and thermal-activation fit used to extract the activation energy.
    \textbf{b}
    Composition dependence of $\rho(T)$ for (Ba$_{1-x}$K$_x$)(Cd$_{0.9}$Mn$_{0.1}$)$_2$As$_2$ (varying $x$; open symbols) and (Ba$_{0.96}$K$_{0.04}$)(Cd$_{1-y}$Mn$_y$)$_2$As$_2$ (varying $y$; solid symbols), illustrating opposite effects of charge (K) doping and spin/disorder (Mn) doping on resistivity.
    \textbf{c}
    Magnetoresistance at 2~K for (Ba$_{0.96}$K$_{0.04}$)(Cd$_{1-y}$Mn$_y$)$_2$As$_2$ with increasing Mn content, defined as $MR=[\rho(H)-\rho(0)]/\rho(0)$, highlighting the strong enhancement toward the heavily Mn-doped regime.
    \textbf{d}
    Field-dependent transport of the heavily Mn-doped composition (Ba$_{0.96}$K$_{0.04}$)(Cd$_{0.7}$Mn$_{0.3}$)$_2$As$_2$: $\rho(T)$ measured under various magnetic fields; inset: ${\rm MR}(H)$ at selected temperatures showing a colossal negative $MR$ approaching $-100\%$ at 2~K with near-saturation at a low field of $\sim$0.35~T.}
    \label{fig:fig4}
\end{figure}

\end{document}